\documentclass[12pt]{jpconf}
\pdfoutput=1
\usepackage{graphicx}
\usepackage{footmisc}
\usepackage{color}
\usepackage{amsmath} 
\usepackage{array,relsize,float}
\usepackage[bitstream-charter]{mathdesign}

\renewcommand{\thefigure}{\arabic{figure}}

\def\beq{\begin{equation}}
\def\eeq{\end{equation}}
\def\bea{\begin{eqnarray}}
\def\eea{\end{eqnarray}}
\def\beqn{\begin{eqnarray}} 
\def\eeqn{\end{eqnarray}}

\renewcommand{\thefootnote}{\fnsymbol{footnote}}

\begin{document}
\title{Precision studies for the partonic kinematics calculation through  Machine Learning}
\author{David F. Renteria-Estrada$^{1}$\footnote[1]{\hspace{1mm}Speaker}}
\address{$^1$ Facultad  de  Ciencias  F\'isico-Matem\'aticas,  Universidad  Aut\'onoma  de  Sinaloa,  Ciudad  Universitaria, CP 80000 Culiac\'an, M\'exico.}
\author{Roger J. Hernandez-Pinto$^{1}$}
\address{$^1$ Facultad  de  Ciencias  F\'isico-Matem\'aticas,  Universidad  Aut\'onoma  de  Sinaloa,  Ciudad  Universitaria, CP 80000 Culiac\'an, M\'exico.}
\author{G. F. R. Sborlini$^{2,3}$}
\address{$^2$ Departamento de F\'isica Fundamental e IUFFyM, Universidad de Salamanca, 
37008 Salamanca, Spain.}
\address{$^3$ Escuela de Ciencias, Ingenier\'ia y Dise\~no, Universidad Europea de Valencia, \\ Paseo de la Alameda 7, 46010 Valencia, Spain.}
\author{Pia Zurita$^{4,5}$}
\address{$^4$ Institut f\"ur Theoretische Physik, Universit\"at Regensburg, 93040 Regensburg, Germany, Universit\"at Regensburg, Germany.}
\address{$^5$ Facultad de Ciencias F\'isicas, Universidad Complutense de Madrid, Pl. Ciencias, 1, 28040 Madrid, Spain.}

\ead{davidrenteria.fcfm@uas.edu.mx}

\begin{abstract}
High Energy collider experiments are moving to the highest precision frontier quickly. The predictions of observables are based on the factorization formula which helps to connect small to large distances. These predictions can be contrasted with experimental measurements and the success of this phenomenological approach is based on the correct description of nature. The application of the method to proton-proton colliders brings new challenges due to the proton structure and the detectors efficiency on reconstructing hadrons. Furthermore, since the non-perturbative distribution functions takes an important role to describe the experimental distributions, the presence of them makes the information of the partons diluted. At Leading Order (LO) in perturbative calculations, the momentum fractions involved in hard scattering processes are known exactly in terms of kinematical variables of initial and final states hadrons. However, at Next-to-Leading Order (NLO) and beyond, a closed analytical formula is not available. Furthermore, from the pure theoretical calculation, the exact definition of the momentum fraction is very challenging. In this work, we report a methodology based on Machine Learning techniques for the extraction of momentum fractions for $p+p\to\pi^++\gamma$ using a Monte Carlo simulation including quantum corrections up to Next-to-Leading Order in Quantum Chromodynamics and Leading Order in Quantum Electrodymics. Our findings point towards a methodology to find the fundamental properties of the internal structure of hadrons because the reconstructed momentum fractions deeply relate our perturbative models with experimental measurements.
\end{abstract}

\setcounter{page}{1}

\section{Introduction}
Technological advances have enabled important breakthroughs in science, and in the case of high energy physics have allowed us to discover new particles and test models. To mention some specific examples, technology has favored perturbative Quantum Chromodynamics (pQCD) to achieve greater precision in computational calculations and to improve the efficiency of such calculations. The trend right now points towards Artificial Intelligence (AI) and Machine Learning (ML) techniques applied to high energy experiments in parton shower Monte Carlo (MC) reconstruction applied to  jets physics\cite{Cranmer:2021gdt}, reconstruction of deep inelastic dispersion kinematics \cite{Diefenthaler:2021rdj,Arratia:2021tsq}, or more phenomenological analysis, including the determination of partonic distribution densities by the NNPDF collaboration \cite{Ball:2008by,Ball:2012cx,DelDebbio:2007ee,Forte:2002fg,NNPDF:2014otw,NNPDF:2019vjt,NNPDF:2021njg,Rojo:2004iq}.

As is well known, the Quantum Chromodynamics (QCD) is a theory that studies the strong interactions between partons, but in hadronic collisions it is necessary to resort to pQCD theory. Calculating the scattering cross-section is a useful measure to study these interactions, so it is important to know the accessible kinematic of the detector how to reconstruct the underlying partonic momentum fractions. In a perturbation theory, these quantities are only accurately determined when analytical calculations are computed up to the Leading Order (LO) precision. However, obtaining the scattering cross-section at LO accuracy is not sufficient to describe precise experimental measurements. To achieve a more reliable the phenomenological description, higher-order calculations are required. In pQCD, this feature happens even for the first or second order, the so called Next-to-Leading Order (NLO) and Next-to-Next-to-Leading Order (NNLO) respectively. Furthermore, the problem grows when we start combining the effects of the interaction of photons with partons and the QCD interactions, since non-perturbative effects must be considered, hadronization.

In this work we propose a methodology based on Neural Networks using Machine Learning (ML) to reconstruct the partonic momentum fractions in photon-hadron production at colliders, including up to NLO QCD and LO QED correction. We base our work on Refs. \cite{PhysRevD.83.074022,Renteria-Estrada:2021zrd}, which we briefly explain in Sec. \ref{sec:detallescomputacionales}. In Sect. \ref{sec:fenomenologia} we report the phenomenological results of studying the hadronic differential cross-section distribution as a function of the phase-space of the system. Finally, in Sec. \ref{sec:resultadosNN} we show the Neural Network (NN) results for predicting the partonic momentum fractions of the scattering process in the initial and final state at LO QED + NLO QCD accuracy. 

\section{Computational details}\label{sec:detallescomputacionales}
The efforts in the scientific community to obtain greater accuracy into quarks and gluons distributions, had contributed to solve the problem of the proton spin crisis. Still, there is a long way to pursue and one direction is the  precise determination of the partonic moments in the pQCD theory. In this context, studying the parton momentum fractions in hadronic collisions constitutes an advance towards the search of greater precision since they carries an imprint of the parton level kinematics. In our study we are interested to perform the MC simulation of the cross-section for the reaction,
\begin{eqnarray}
	p\,p\to \pi^+ + \gamma\,,
\end{eqnarray}
where the photon shall be produced by the direct interaction of the original partons in the collision, which we will refer to as the hard photon. Fortunately, the dispersion between hadrons can be extracted by means of hard process factorization properties in the QCD \cite{Collins:1989gx}. Therefore, the differential cross-section of the proton-proton process is obtained through the convolution of the Parton Distribution Function (PDF), Fragmentation Function (FF) and the differential partonic cross-section. The non-perturbative part corresponds to the PDF and the FF, which provide us the information on the probabilities of finding the partons inside the colliding hadrons and the probability of generating a given meson/hadron from a certain parton, respectively. Explicitly the differential cross-section of producing a positively charged pion in addition with a photon in the proton-proton collisions is given by,
\begin{align}
	\nonumber d\sigma_{p_1 \, p_2 \to \pi \, \gamma} =& \sum_{a_1 a_2 a_3 a_4} \int dx_1 dx_2 dz dz'\, f^{(p_1)}_{a_1}(x_1,\mu_I)
	f^{(p_2)}_{a_2}(x_2,\mu_I) \, d^{(\pi)}_{a_3}(z,\mu_F) d^{(\gamma)}_{a_4}(z',\mu_F)  \nonumber\\
	&\times d\hat\sigma_{a_1\,a_2 \to a_3 \, a_4}(x_1,x_2,z,z',\mu_R) \, , 
	\label{eq:cs}
\end{align}
where we sum over all flavors of quarks and gluons $a_1,\, a_2\,$ and $a_3$, $f^{(h)}_{a}(x,\mu_I)$ represents the PDF which indicates the probability of finding the parton $a$ inside the hadron $h$ with momentum fraction $x$ and initial energy scale $\mu_I$. Analogously $d^{(h)}_{b}(z,\mu_F)$ stands for the probability that the parton $b$  produced during the hard scattering, hadronizes in the hadron $h$ with momentum fraction $z$ at a typical energy scale $\mu_F$. Finally, we cast in $d\hat\sigma_{a_1\,a_2 \to a_3 \, a_4}$ the differential partonic cross-section of the interaction between the partons $a_1$ and $a_2$ producing the partons $a_3$ and $a_4$. In practice, the nature of the detected photons is indistinguishable, however, there are reconstructions algorithms that permit to retain a high proportion of photons expected to come directly from the parton-level interaction (i.e. from the hard-scattering). For this purpose, we impose a criteria to isolate photons\cite{Frixione:1998jh} which allows us to obtain independence of the parton-to-photon fragmentation function \footnote{The smooth isolation prescription algorithm can be found in Refs. \cite{renteria2021analysis,Frixione:1998jh}.} and also to solve the collinear divergences. Therefore, the differential cross-section of producing a positively charged pion plus a hard photon into proton-proton collision can be rewritten as follows, 
\begin{align}
	d\sigma_{p_1 \, p_2 \to \pi \, \gamma} = &\sum_{a_1 a_2 a_3} \int dx_1 dx_2 dz \, f^{(p_1)}_{a_1}(x_1,\mu_I)
	f^{(p_2)}_{a_2}(x_2,\mu_I) \, d^{(\pi)}_{a_3}(z,\mu_F) \nonumber\\
	&\times d\hat\sigma_{a_1\,a_2 \to a_3 \, \gamma}^{\rm{ISO}}(x_1,x_2,z,\mu_R) \label{eq:csISO}
\end{align}
where $d\hat\sigma_{a_1\,a_2 \to a_3 \, \gamma}^{\rm{ISO}}$ is the differential cross-section to generate a parton $a_3$ and an isolated photon from the interaction between partons $a_1$ and $a_2$ taking into account the isolation algorithm. In our case, we perform the calculation considering including up to LO QED + NLO QCD. In this way, the partonic cross-section is the sum of the QED plus the QCD contributions, namely,
\begin{eqnarray}
	d\hat\sigma_{a_1\,a_2 \to a_3 \, \gamma}^{\rm{ISO}} = d\hat\sigma_{a_1\,a_2 \to a_3 \, \gamma}^{\rm{ISO,\, QED}} + d\hat\sigma_{a_1\,a_2 \to a_3 \, \gamma}^{\rm{ISO,\,QCD}}\, .
	\label{eq:csQED+QCD}
\end{eqnarray}
Once we have applied the isolation criteria, the partonic channels for the 2 $\to$ 2 processes in QCD are, 
\begin{eqnarray}
	{q\bar{q}} \to {\gamma g}, \quad {qg \to \gamma q}\, , 
\end{eqnarray}
meanwhile, the channels in the QED interaction are, 
\begin{eqnarray}
	{q\bar{q}} \to {\gamma \gamma}, \quad {q\gamma \to \gamma q}\, , 
\end{eqnarray}
and, finally, the processes 2$\to$ 3 considered in the NLO contribution of QCD are those as,  
\begin{eqnarray}
	&{q\bar{q}} \to {\gamma g g}\, , \quad {qg \to \gamma g q}\, , \quad {gg \to \gamma q \bar{q}}\, , & \nonumber\\
	&{q\bar{q} \to \gamma Q \bar{Q}}\, , \quad {qQ \to \gamma q Q} \, .&
\end{eqnarray}

\section{Phenomenological results}\label{sec:fenomenologia} 
The calculations described in Sec. \ref{sec:detallescomputacionales} were implemented in the MC integrator based in Ref. \cite{PhysRevD.83.074022} and considered two different scenarios for the scattering kinematics. In the first scenario, we seek to study the kinematic configuration of PHENIX in the RHIC experiment with a center-of-mass (c.m.) energy of $\sqrt{S_{\rm CM}}=500$ GeV, and considering the phenomenological cuts,  
\begin{eqnarray}
	&\{|\eta^\pi|,\,|\eta^\gamma|\}\leq 0.35 \, ,& \nonumber \\
	&p_T^\pi \geq 2\, {\rm GeV}\, , &\nonumber \\ 
	&5\,{\rm GeV} \leq p_T^\gamma \leq 15\, {\rm GeV}\, ,&  
\end{eqnarray}
where $\eta$ and $p_T$ are the pseudorapidity and transverse momentum, respectively. On the other hand, in the second scenario we explore a c.m. energy of $\sqrt{S_{\rm CM}}=13$ TeV corresponding to LHC RUN II configuration, keeping the same cuts for the pion kinematics. However, motivated by studies about the photon detection efficiency in ATLAS and CMS different energies \cite{Kontaxakis:2018xiw,ATLAS:2019dpa,ATLAS:2019qmc}, we impose the following cuts on the photon kinematics, 
\begin{eqnarray}
	&|\eta^\gamma| \leq 2.5 \, ,& \nonumber \\
	&p_T^\gamma \geq 30\, {\rm GeV}\, .&  
\end{eqnarray}
In addition, it is important to mention that we implement the restriction $|\phi^\pi-\phi^\gamma|\geq 2$ on the azimuthal angles of the pion and photon, in order to keep a {\it back-to-back} configuration. Finally, we set the renormalization and factorization scales as the average of the pion and photon transverse momenta,
\begin{eqnarray}
	\mu_R = \mu_I = \mu_F = \frac{p_T^\pi+p_T^\gamma}{2}\, .
\end{eqnarray}

\subsection{Partonic kinematics analysis}\label{sec:analisiscinematica}
\begin{figure}[H]
	\begin{center}
	\includegraphics[width=0.49\textwidth]{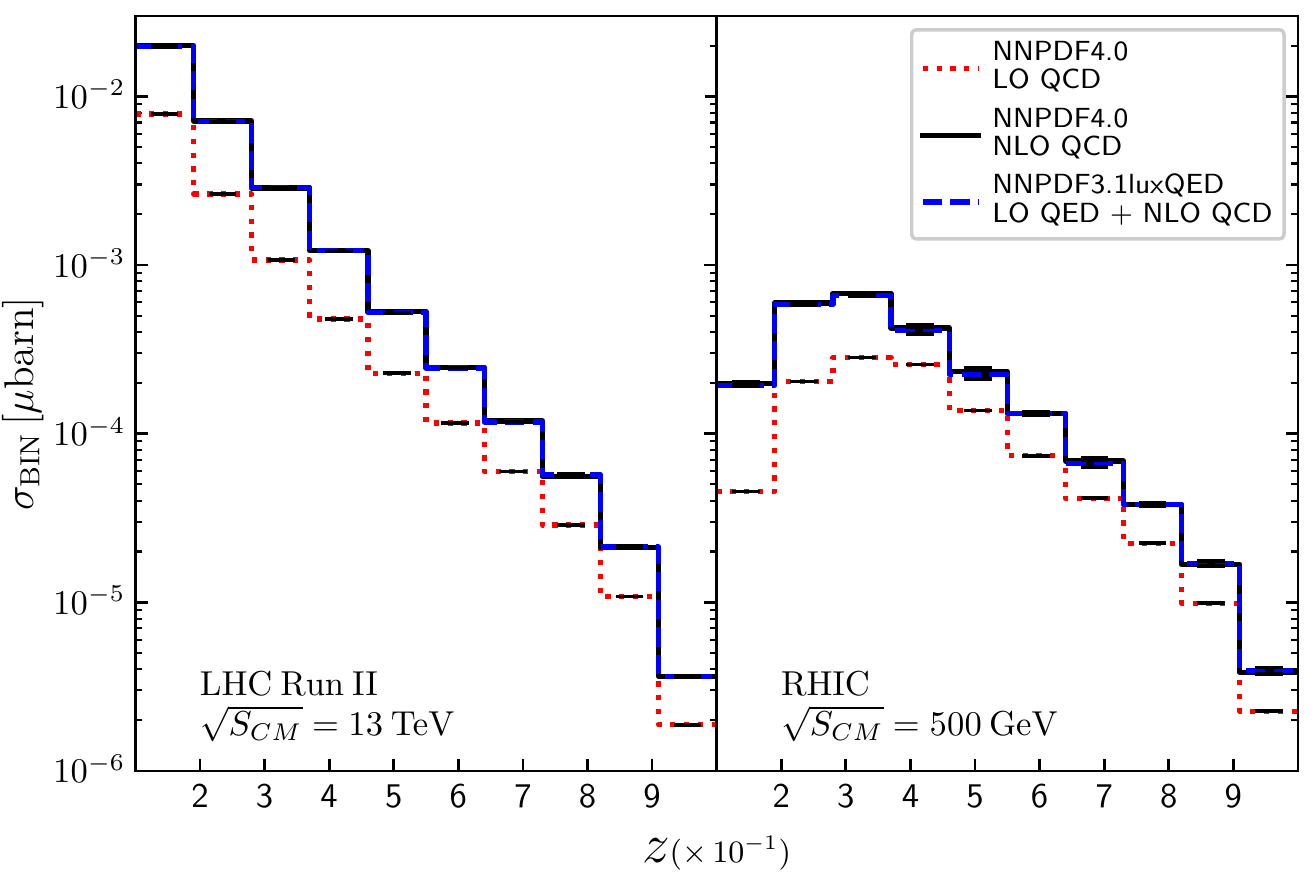}
	\caption{Cross-section as a function of the partonic momentum fractions, for RHIC  and LHC Run II . 
	}
    \label{fig:ONEdimensionZ} 
	\end{center}
\end{figure}

Now, we study the momentum fraction distributions with the purpose of imposing constraints on the NN training. As it is well known, the momentum fractions are exactly determined at LO accuracy, explicitly as follows,
\begin{align}
	x_{1,2} &= \frac{p_T^\gamma\left(\exp{(\eta^{\pm\pi})}+\exp{(\eta^{\pm\gamma})}\right)}{\sqrt{S_{CM}}}\label{eq:xLO}\, ,\\
	z &= \frac{p_T^\pi}{p_T^\gamma}\label{eq:zLO}\, .
\end{align}
To compute Eq. (\ref{eq:csISO}), we use in the MC simulation the PDF set {\tt NNPDF4.0NLO} \cite{NNPDF:2021njg} for the NLO QCD prediction, while for LO QED + NLO QCD quantum corresctions we used the set {\tt NNPDF3.1luxQEDNLO} \cite{Campbell:2018wfu,Bertone:2017bme,Manohar:2016nzj,Manohar:2017eqh}. Likewise, we selected the most up-to-date FF of pions at NLO given by collaboration {\tt DSS2014} \cite{deFlorian:2014xna,deFlorian:2007ekg}. Additionally, by simplicity we define the same partonic momentum fraction for both protons as $x=\{x_1,\, x_2\}$. 
\begin{figure}[H]
	\begin{center}
		\includegraphics[width=0.49\textwidth]{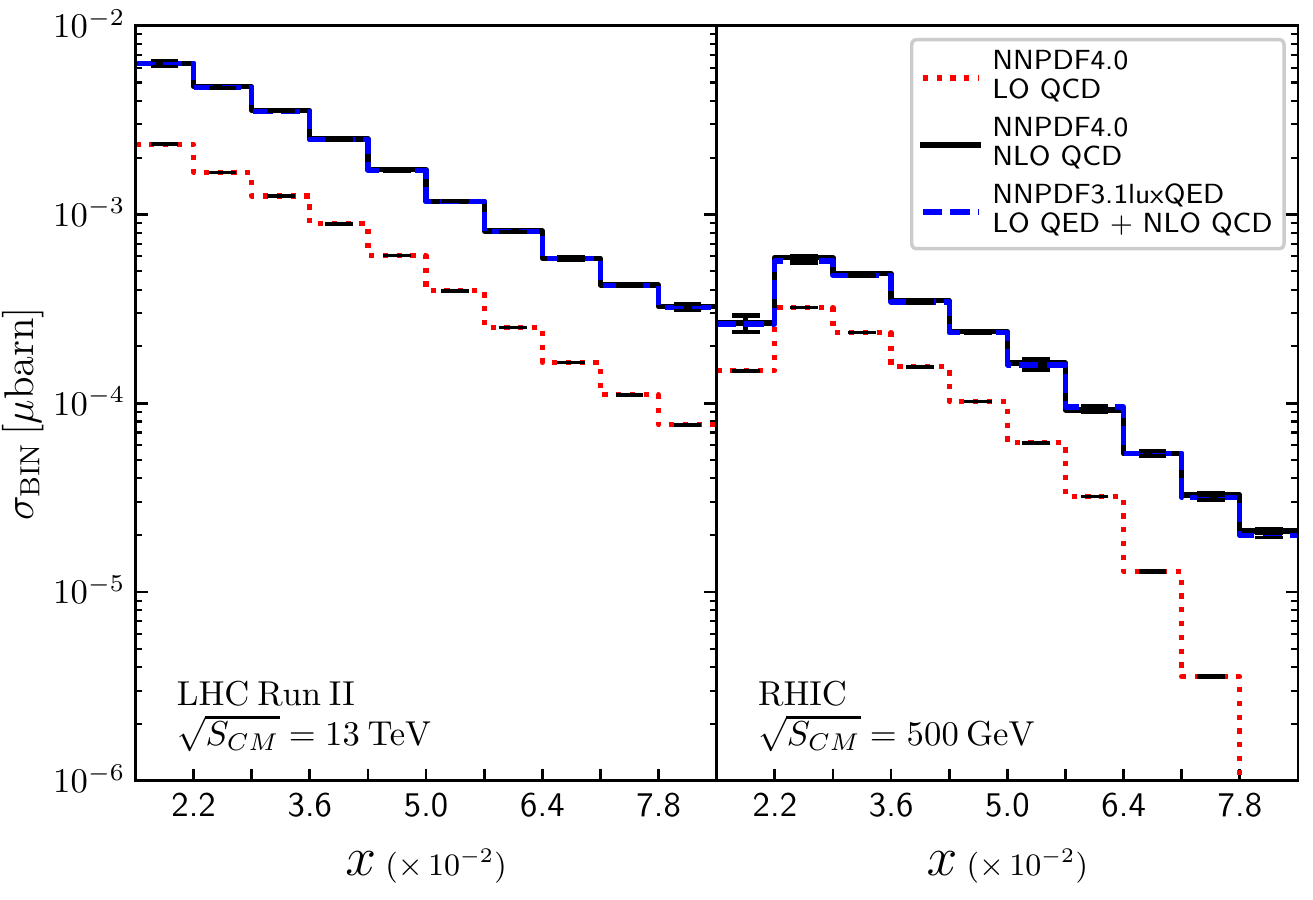}
		\caption{Cross-section as a function of the partonic momentum fractions, for RHIC  and LHC Run II . 
		}
	\label{fig:ONEdimensionX} 
	\end{center}
\end{figure}
In Fig. \ref{fig:ONEdimensionX} we display the differential cross-section per bin of the $pp\to\pi^++\gamma$ scattering process as a function of the partonic momentum fraction $x$, both for energies of RHIC (right) and LHC RUN II (left) experiments. A significant contribution to higher-order corrections is observed with respect to the LO contribution in QCD (red dots).  The effects of QED and QCD (blue dashed line) are smaller with respect to only NLO QCD contribution (black line). Furthermore, as expected from LO, the maximum limit imposed on the photon momentum $p_T^\gamma \leq 15$ GeV at RHIC energies, manifests itself in a peak of the $d\sigma_{pp\to\pi+\gamma}$ distribution at $x\approx 2.5\times 10^{-2}$, both for the Born level and for higher-orders. Similarly, in Eq. (\ref{eq:xLO}) the $p_T$ and $\eta$ restrictions in RHIC, limit the range that $x$ can be computed to $x_{{\rm max}}\approx 0.01$.  For this reason, we limited our analysis to observing the LHC kinematics in the same RHIC range. As a matter of fact, in Fig. \ref{fig:ONEdimensionZ} we show an analysis of the differential cross-section distribution analogous to Fig. \ref{fig:ONEdimensionX} but as a function of $z$. Here, we can see that the LHC kinematics implies that cross-section decreases faster than the calculated with RHIC kinematics. The distributions show a peak at $z_{\rm peak}\approx 0.35$ in RHIC whereas $z_{\rm peak}\approx 0.25$ for the LHC RUN II. It is important to mention that those peaks correspond to a greater extent to the constraints implemented in the FF. 
\subsection{NN prediction}\label{sec:resultadosNN}

\begin{table*}[b]
	\renewcommand{\arraystretch}{1} %
	\centering
	\begin{tabular}{@{\extracolsep{0.0 cm}}|c|c|c|c|c|}
		\hline \hline
		&\small{$X_{{\rm REC}}$ (LO)} & \small{$Z_{{\rm REC}}$ (LO)} & \small{$X_{{\rm REC}}$ (NLO)} & \small{$Z_{{\rm REC}}$ (NLO)}\\
		\hline
		$\#$ of hidden layers & \small{2}& \small{1} &\small{5} & \small{5} \\    
		$\#$ of neurons/layer & \small{200}& \small{100} &\small{300} & \small{300}\\
		activation function & ReLU& ReLU& ReLU& ReLU\\
		$\#$ iterations & $1\times\,10^{5}$ & $1\times\,10^{5}$ & $1\times\,10^{12}$ & $1\times\,10^{12}$ \\
		learning rate & $1\times\,10^{-3}$& $1\times\,10^{-3}$& $1\times\,10^{-4}$& $1\times\,10^{-4}$ \\
		\hline \hline 
	\end{tabular}
	\caption{Architecture for the MLP best fit parameters for the reconstruction of the momentum fractions at LO in QCD: $X_{\rm REC}$(LO) and $Z_{\rm REC}$(LO) (second and third columns), and for the momentum fractions at NLO QCD + LO QED: $X_{\rm REC}$(NLO) and $Z_{\rm REC}$(NLO) (fourth and fifth columns).} 
	\label{tab:paramMLP}
\end{table*} 
Taking into account the effects of kinematical cuts on the cross-section distribution, we will focus our efforts on reconstructing the partonic momentum fractions $x$ and $z$. Since we now know that in the LO case our phase-space is well defined, we will use this information to check the reliability of our method. In this task, we implement a NN with Python 3 library {\tt scikit-learn} \cite{Pedregosa:2011ork}, specifically, we train a Multilayer Perceptron (MLP) whose architecture is shown in Tab.~\ref{tab:paramMLP}. For this purpose, we define $X_{\rm REC}$ and $Z_{\rm REC}$ as target variables of the NN. For the training to predict the $x$ variable in the LO case, we set the MLP with two hidden layers and 200 neurons per layer, using $10^{5}$ interactions per event number. In addition, we find some parameters sensitive to the output variables, such as the maximum number of interactions and the learning rate. Finally, for all architectures we use the Rectified Linear Unit (ReLU) activation function.

\begin{figure}[H]
	\begin{center}
		\includegraphics[width=0.49\textwidth]{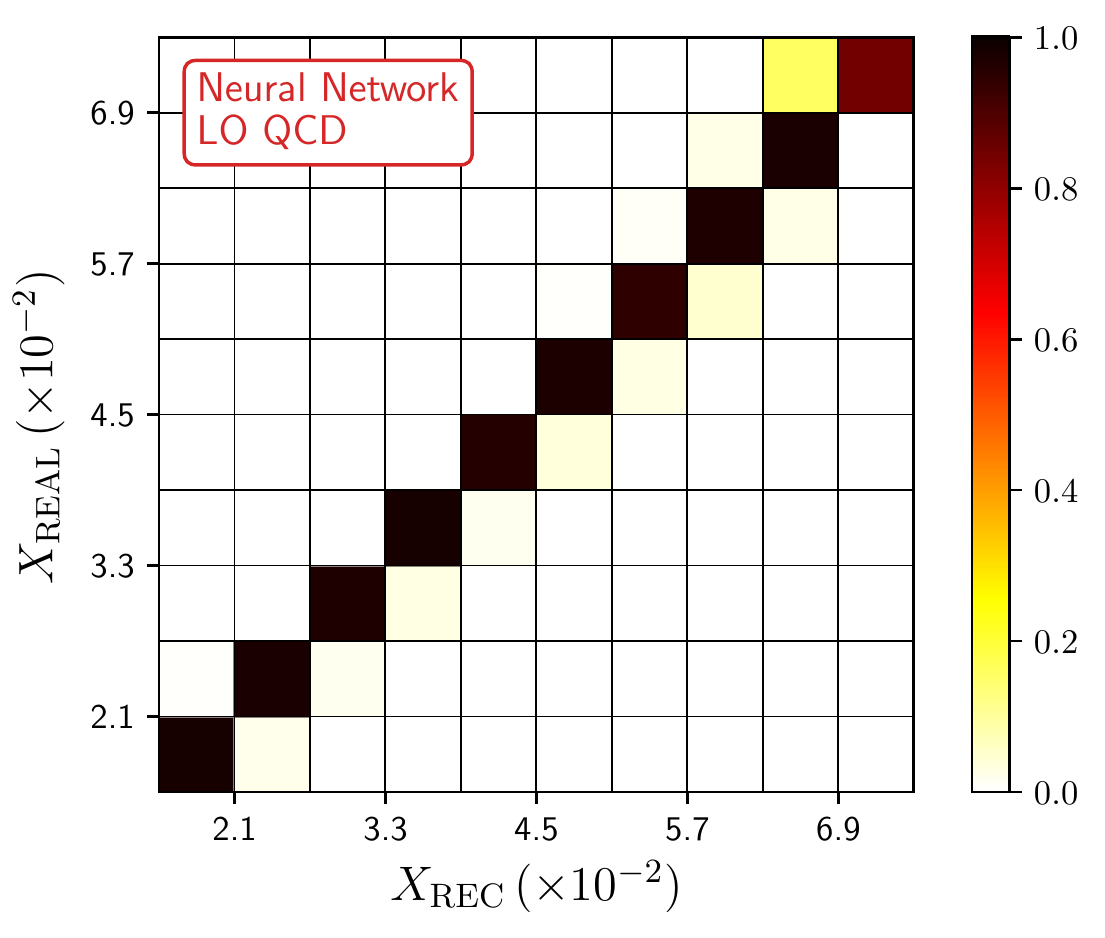}
		\includegraphics[width=0.49\textwidth]{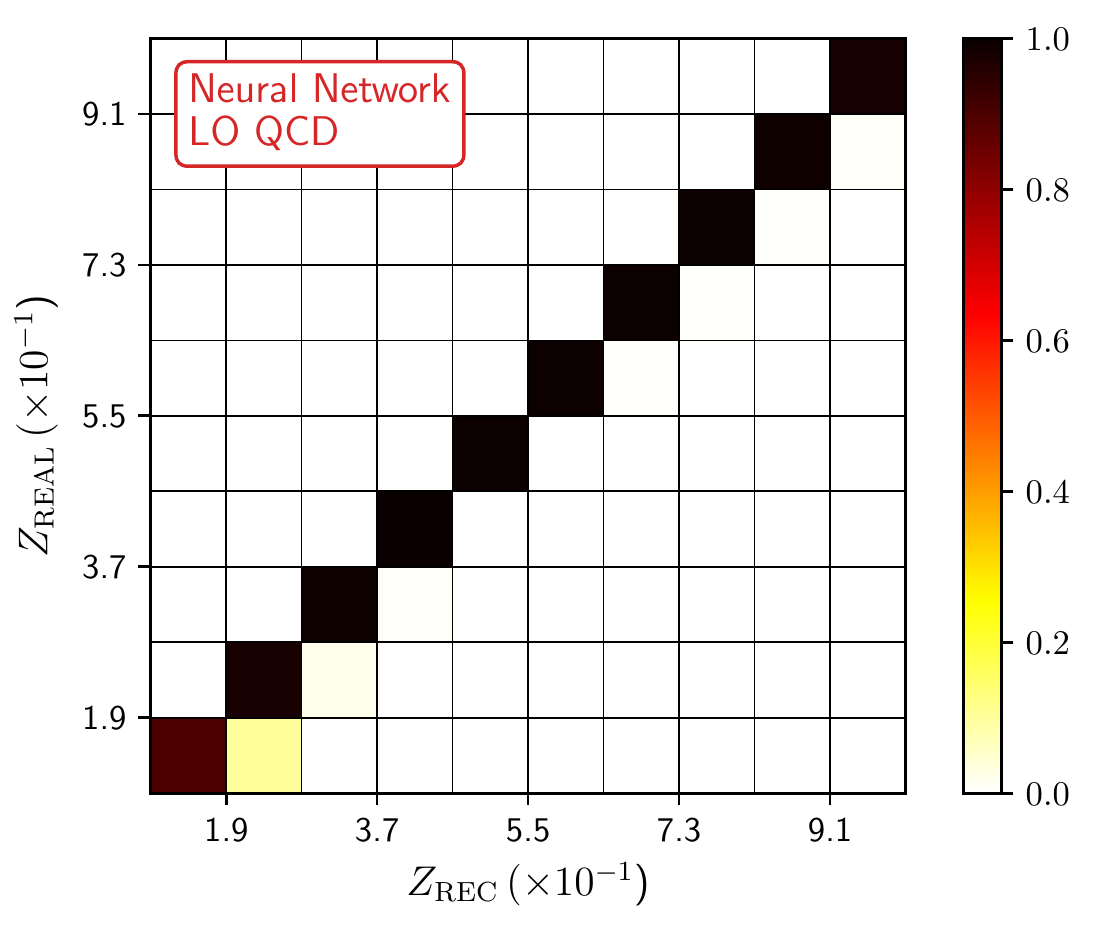}
		\caption{Comparison of the momentum fractions at LO accuracy obtained with MLP neural networks with the parameters given in Table \ref{tab:paramMLP}.}
		\label{fig:LOMLP}
	\end{center}
\end{figure}

The methodology to train the NN consists of generating random points ($x,\,z$), assigning them as the target variables. In this way, we use 20\% of the data to test the results of the NN and 80\% for training the NN. In the LO case, the inputs values are the kinematic variables accessible in the experimental environment, this mean,
\begin{eqnarray}
	{\cal V}_{\rm Exp} = \{p_T^\gamma,p_T^\pi,\eta^\gamma,\eta^\pi,\cos(\phi^\pi-\phi^\gamma)\} \, .
	\label{eq:VARIABLES}
\end{eqnarray}
Conversely, due to the discussion in Sec. \ref{sec:analisiscinematica}, the kinematic cutoffs  imposed restrict the range in which we can measure $x$ and $z$, therefore, we reconstruct the partonic momentum fractions in the range $x\in(0.015,\, 0.075)$ and $z\in(0.1,\, 1)$. The results at LO are displayed in Fig. \ref{fig:LOMLP}, where we have defined the variables $X_{\rm REAL}$ and $Z_{\rm REAL}$ as the variables randomly generated by our MC simulation. The color scale indicates the probability that the Neural Network predicts $\{X_{\rm REC},\,Z_{\rm REC}\}$ in the same bin as $\{X_{\rm REAL},\,Z_{\rm REAL}\}$. With this architecture, the NN can access the momentum fractions with probability greater than 90\% for $x<6.9\times 10^{-2}$, and a probability around 80\% for $x>6.9\times 10^{-2}$. Meanwhile, the NN trained to predict $z$ fraction, have simpler architecture and reconstructs the momentum fraction with  success rate close to 100\% in $z>0.19$. At the Born level, $x$ is proportional to the photon transverse momentum, while $z$ is inversely proportional, so the imposed upper bound of $P_T^{\gamma}\leq 15\,{\rm GeV}$ has repercussions on the reconstruction of $x$ when large values of $X_{{\rm REAL}}$ and low values of $Z_{\rm REAL}$ are explored.

Once we have checked the advantage of the NN, we train it to perform predictions at NLO level. However, since having a better fit becomes more complex, we opt for a more robust architecture. By the nature of the process, the experiment is not able to distinguish between a LO and NLO contribution, therefore, for the NLO QCD + LO QED correction it is necessary to discretize the kinematic variables used as input values namely, 
\begin{eqnarray}
	p = \{{\bar p}_T^\gamma,{\bar p}_T^\pi,{\bar\eta}^\gamma,{\bar\eta}^\pi,{\overline{\cos}}(\phi^\pi-\phi^\gamma)\}\in{\cal \bar{V}}_{\rm Exp} \, ,
	\label{eq:VARIABLESpromedio}
\end{eqnarray}
where $\bar{a}$ denotes the mean value of each variable of ${\cal V}_{\rm Exp}$ per bin. Now we discretize the space with ten bins for $p_T$, five bins for $\eta$ and six bins for $\cos{(\phi^\pi-\phi^\gamma)}$. This leads us to weigh the partonic momentum fractions with the average cross-section corresponding to the bin of the event ${\cal{\bar V}}_{\rm Exp}$. Explicitly, the target variables are now, 
\begin{eqnarray}
	X_{1,\rm REC} &=& \sum_i \, (x_1)_i \frac{d \sigma_j}{d x_1} (p_j;(x_1)_i) \, ,\label{eq:x1BINj}\\ 
	X_{2,\rm REC} &=& \sum_i \, (x_2)_i \frac{d \sigma_j}{d x_2} (p_j;(x_2)_i) \, ,\label{eq:x2BINj}\\
	Z_{\rm REC}  &=& \sum_i \, z_i \frac{d \sigma_j}{d z} (p_j;z_i) \, ,\label{eq:zBINj}
\end{eqnarray}

\begin{figure}[H]
	\begin{center}
	\includegraphics[width=0.49\textwidth]{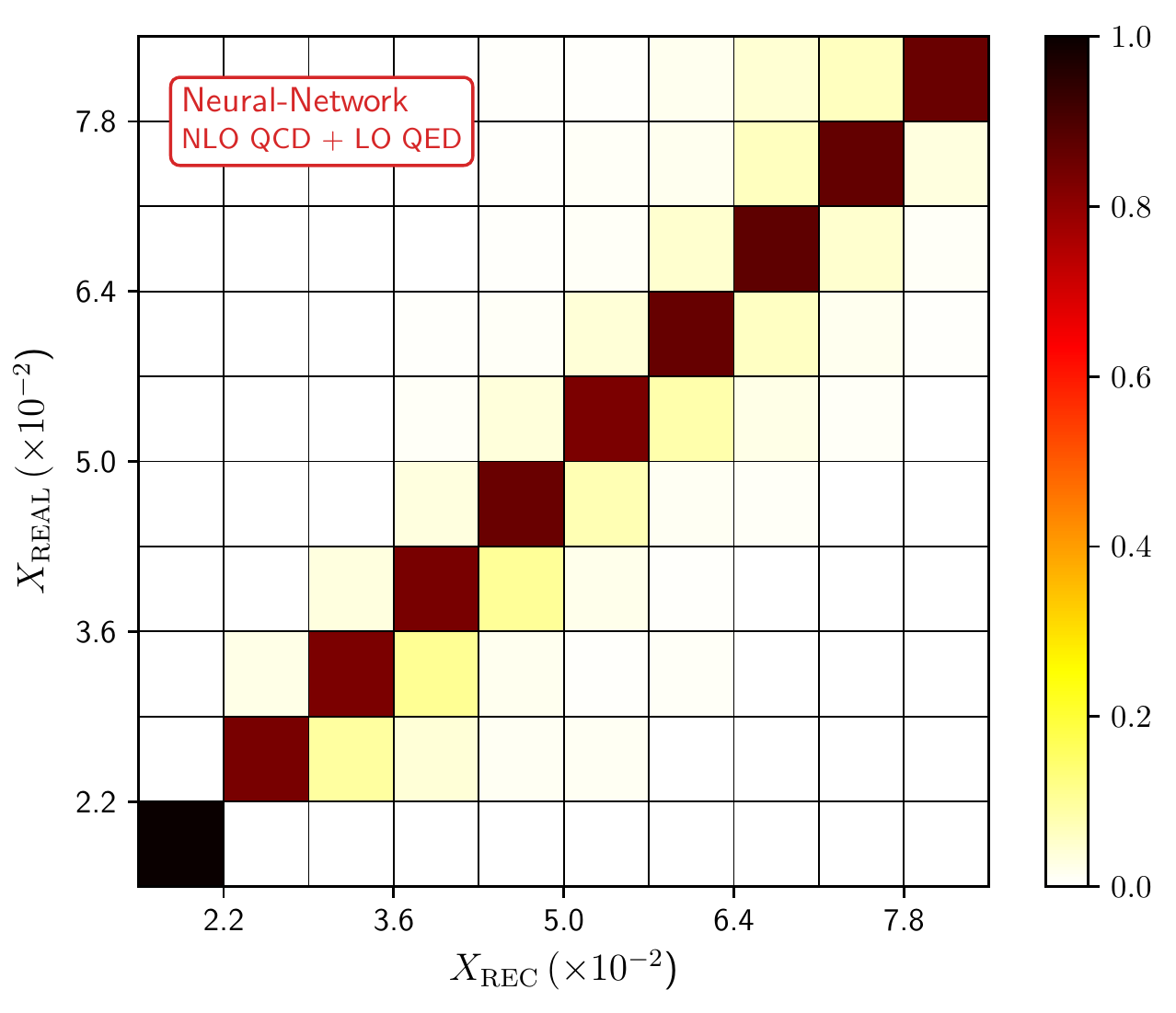}
	\includegraphics[width=0.49\textwidth]{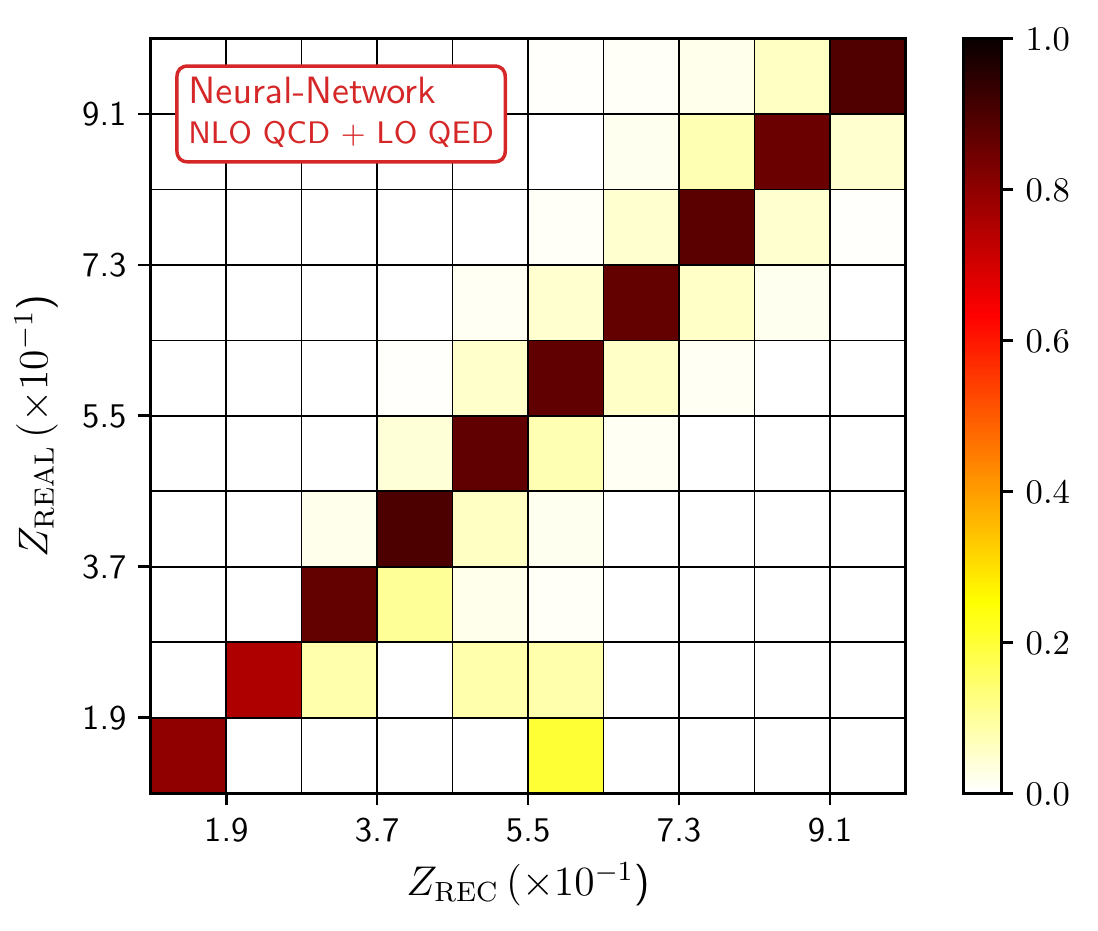}
	\caption{Comparison of the momentum fractions at LO QED + NLO QCD accuracy obtained with MLP neural networks with the parameters given in Table \ref{tab:paramMLP}.} 
	\label{fig:NLOMLP}
	\end{center}
\end{figure}

The results at NLO QCD + LO QED accuracy are presented in Fig. \ref{fig:NLOMLP}, for which we use a MLP whose parameters are those of Tab. \ref{tab:paramMLP} (fourth and fifth columns). We generated $10^9$ events for LO case, from which only 30\% passed the kinematic cuts. In addition, the discretization of the input and output values at NLO led to generate 15000 variables in our phase-space. For both $x$ and $z$, the NN showed favorable results and a remarked improvement over the work made in Ref. \cite{PhysRevD.83.074022}. The prediction for $x$ is excellent in the lowest rank bin, while in the rest bins the probability of reconstruction oscillates around 80\%. On the other hand, the fraction $z$ achieves the best prediction in the range $z\in(0.28,1)$; on the other hand, we find that for low values of $z$, our result continues to show a strong correlation.
\section{Conclusions and outlook}\label{sec:conclusiones}
In the context of hadron collisions, it is difficult to obtain closed analytical formulae to related parton momenta fraction and experimentally-accessible variables. This fact leads us to use only approximate formulas in the calculation of the scattering cross-section. In this work, we seek to access the partonic momentum fractions for the production of a hadron plus a direct photon in proton-proton collisions using ML tools. First, we calculated the cross-section differential distribution of the scattering process $pp\to \pi^++\gamma$ with RHIC and LHC RUN II kinematics.  The results indicated that the kinematical cuts imposed by both experiments limit the range of accessibility of $x$. On the other hand,  we noted that RHIC experiment has a peak in both $x$ and $z$, which was observed due to restrictions of $p_T^\gamma$ and pion fragmentation. Likewise, we have built a MLP NN to access the partonic momentum fractions at NLO QCD + LO QED accuracy. In the first instance, we obtained predictions with LO QCD accuracy to compare the fitting efficiency of the NN, and we obtained nearly perfect results. Subsequently, with the help of Deep Learning, we accessed the partonic momentum fractions with NLO QCD + LO QED precision getting results with higher accuracy. Finally, this work not only lays the foundation for higher accuracy predictions of High Energy Physics, but also, as mentioned in Ref. \cite{Ochoa-Oregon:2023ktx}, it can help to find constraints on the final states of heavy hadrons.

\section*{Acknowledgements} 
The work of D. F. R.-E. and R. J. H.-P. is supported by CONACyT (M\'exico) through the Project No. A1- S-33202 (Ciencia B\'asica) and Ciencia de Frontera 2021-2042; in addition by PROFAPI 2022 Grant No. PRO\_A1\_024 (Universidad Aut\'onoma de Sinaloa). Besides, R. J. H.-P. is also funded by Sistema Nacional de Investigadores from CONACyT. P.Z. acknowledges support from the Deutsche Forschungsgemeinschaft (DFG, German Research Foundation) - Research Unit FOR 2926, grant number 430915485. GS was partially supported by EU Horizon 2020 research and innovation program STRONG-2020 project under grant agreement No. 824093 and H2020-MSCA-COFUND-2020 USAL4EXCELLENCE-PROOPI-391 project under grant agreement No 101034371."
\section{References}
\bibliographystyle{JHEP}

\bibliography{ref}

\providecommand{\href}[2]{#2}\begingroup\raggedright\begin{thebibliography}{10}

\bibitem{Cranmer:2021gdt}
K.~Cranmer, M.~Drnevich, S.~Macaluso and D.~Pappadopulo, \emph{{Reframing Jet
  Physics with New Computational Methods}},
  \href{http://dx.doi.org/https://doi.org/10.1051/epjconf/202125103059}{\emph{EPJ
  Web Conf.} {\bf 251} (2021) 03059},
  [\href{http://arxiv.org/abs/2105.10512}{{\tt 2105.10512}}].

\bibitem{Diefenthaler:2021rdj}
M.~Diefenthaler, A.~Farhat, A.~Verbytskyi and Y.~Xu, \emph{{Deeply learning
  deep inelastic scattering kinematics}},
  \href{http://dx.doi.org/https://doi.org/10.1140/epjc/s10052-022-10964-z}{\emph{Eur.
  Phys. J. C} {\bf 82} (2022) 1064},
  [\href{http://arxiv.org/abs/2108.11638}{{\tt 2108.11638}}].

\bibitem{Arratia:2021tsq}
M.~Arratia, D.~Britzger, O.~Long and B.~Nachman, \emph{Reconstructing the
  kinematics of deep inelastic scattering with deep learning},
  \href{http://dx.doi.org/https://doi.org/10.1016/j.nima.2021.166164}{\emph{Nucl.
  Instrum. Meth. A} {\bf 1025} (2022) 166164},
  [\href{http://arxiv.org/abs/2110.05505}{{\tt 2110.05505}}].

\bibitem{Ball:2008by}
{\scshape NNPDF} collaboration, R.~D. Ball, L.~Del~Debbio, S.~Forte,
  A.~Guffanti, J.~I. Latorre, A.~Piccione et~al., \emph{{A Determination of
  parton distributions with faithful uncertainty estimation}},
  \href{http://dx.doi.org/https://doi.org/10.1016/j.nuclphysb.2008.09.037}{\emph{Nucl.
  Phys. B} {\bf 809} (2009) 1--63}, [\href{http://arxiv.org/abs/0808.1231}{{\tt
  0808.1231}}].

\bibitem{Ball:2012cx}
R.~D. Ball et~al., \emph{{Parton distributions with LHC data}},
  \href{http://dx.doi.org/https://doi.org/10.1016/j.nuclphysb.2012.10.003}{\emph{Nucl.
  Phys. B} {\bf 867} (2013) 244--289},
  [\href{http://arxiv.org/abs/1207.1303}{{\tt 1207.1303}}].

\bibitem{DelDebbio:2007ee}
{\scshape NNPDF} collaboration, L.~Del~Debbio, S.~Forte, J.~I. Latorre,
  A.~Piccione and J.~Rojo, \emph{{Neural network determination of parton
  distributions: The Nonsinglet case}},
  \href{http://dx.doi.org/https://doi.org/10.1088/1126-6708/2007/03/039}{\emph{JHEP}
  {\bf 03} (2007) 039}, [\href{http://arxiv.org/abs/hep-ph/0701127}{{\tt
  hep-ph/0701127}}].

\bibitem{Forte:2002fg}
S.~Forte, L.~Garrido, J.~I. Latorre and A.~Piccione, \emph{{Neural network
  parametrization of deep inelastic structure functions}},
  \href{http://dx.doi.org/https://doi.org/10.1088/1126-6708/2002/05/062}{\emph{JHEP}
  {\bf 05} (2002) 062}, [\href{http://arxiv.org/abs/hep-ph/0204232}{{\tt
  hep-ph/0204232}}].

\bibitem{NNPDF:2014otw}
{\scshape NNPDF} collaboration, R.~D. Ball et~al., \emph{{Parton distributions
  for the LHC Run II}},
  \href{http://dx.doi.org/https://doi.org/10.1007/JHEP04(2015)040}{\emph{JHEP}
  {\bf 04} (2015) 040}, [\href{http://arxiv.org/abs/1410.8849}{{\tt
  1410.8849}}].

\bibitem{NNPDF:2019vjt}
{\scshape NNPDF} collaboration, R.~Abdul~Khalek et~al., \emph{{A first
  determination of parton distributions with theoretical uncertainties}},
  \href{http://dx.doi.org/https://doi.org/10.1140/epjc/s10052-019-7364-5}{\emph{Eur.
  Phys. J.} {\bf C} (2019) 79:838},
  [\href{http://arxiv.org/abs/1905.04311}{{\tt 1905.04311}}].

\bibitem{NNPDF:2021njg}
{\scshape NNPDF} collaboration, R.~D. Ball et~al., \emph{{The path to proton
  structure at 1\% accuracy}},
  \href{http://dx.doi.org/https://doi.org/10.1140/epjc/s10052-022-10328-7}{\emph{Eur.
  Phys. J. C} {\bf 82} (2022) 428},
  [\href{http://arxiv.org/abs/2109.02653}{{\tt 2109.02653}}].

\bibitem{Rojo:2004iq}
J.~Rojo and J.~I. Latorre, \emph{{Neural network parametrization of spectral
  functions from hadronic tau decays and determination of QCD vacuum
  condensates}},
  \href{http://dx.doi.org/https://doi.org/10.1088/1126-6708/2004/01/055}{\emph{JHEP}
  {\bf 01} (2004) 055}, [\href{http://arxiv.org/abs/hep-ph/0401047}{{\tt
  hep-ph/0401047}}].

\bibitem{PhysRevD.83.074022}
D.~de~Florian and G.~F.~R. Sborlini, \emph{Hadron plus photon production in
  polarized hadronic collisions at next-to-leading order accuracy},
  \href{http://dx.doi.org/https://10.1103/PhysRevD.83.074022}{\emph{Phys. Rev.
  D} {\bf 83} (Apr, 2011) 074022}.

\bibitem{Renteria-Estrada:2021zrd}
D.~F. Renter\'\i{}a-Estrada, R.~J. Hern\'andez-Pinto, G.~F.~R. Sborlini and
  P.~Zurita, \emph{{Reconstructing partonic kinematics at colliders with
  machine learning}},
  \href{http://dx.doi.org/https://10.21468/SciPostPhysCore.5.4.049}{\emph{SciPost
  Phys. Core} {\bf 5} (2022) 049}, [\href{http://arxiv.org/abs/2112.05043}{{\tt
  2112.05043}}].

\bibitem{Collins:1989gx}
J.~C. Collins, D.~E. Soper and G.~F. Sterman, \emph{{Factorization of Hard
  Processes in QCD}},
  \href{http://dx.doi.org/https://10.1142/9789814503266_0001}{\emph{Adv. Ser.
  Direct. High Energy Phys.} {\bf 5} (1989) 1--91},
  [\href{http://arxiv.org/abs/hep-ph/0409313}{{\tt hep-ph/0409313}}].

\bibitem{Frixione:1998jh}
S.~Frixione, \emph{{Isolated photons in perturbative QCD}},
  \href{http://dx.doi.org/https://10.1016/S0370-2693(98)00454-7}{\emph{Phys.
  Lett. B} {\bf 429} (1998) 369--374},
  [\href{http://arxiv.org/abs/hep-ph/9801442}{{\tt hep-ph/9801442}}].

\bibitem{renteria2021analysis}
D.~F. Renter{\'\i}a-Estrada, R.~J. Hern{\'a}ndez-Pinto and G.~Sborlini,
  \emph{Analysis of the internal structure of hadrons using direct photon
  production},
  \href{http://dx.doi.org/https://doi.org/10.3390/sym13060942}{\emph{Symmetry}
  {\bf 13} (2021) 942}.

\bibitem{Kontaxakis:2018xiw}
{\scshape CMS} collaboration, P.~Kontaxakis, \emph{{The Level-1 CMS electron
  and photon trigger for the LHC Run II}},
  \href{http://dx.doi.org/https://10.22323/1.321.0073}{\emph{PoS} {\bf
  LHCP2018} (2018) 073}.

\bibitem{ATLAS:2019dpa}
{\scshape ATLAS} collaboration, G.~Aad et~al., \emph{{Performance of electron
  and photon triggers in ATLAS during LHC Run 2}},
  \href{http://dx.doi.org/https://10.1140/epjc/s10052-019-7500-2}{\emph{Eur.
  Phys. J. C} {\bf 80} (2020) 47}, [\href{http://arxiv.org/abs/1909.00761}{{\tt
  1909.00761}}].

\bibitem{ATLAS:2019qmc}
{\scshape ATLAS} collaboration, G.~Aad et~al., \emph{{Electron and photon
  performance measurements with the ATLAS detector using the
  2015\textendash{}2017 LHC proton-proton collision data}},
  \href{http://dx.doi.org/https://10.1088/1748-0221/14/12/P12006}{\emph{JINST}
  {\bf 14} (2019) P12006}, [\href{http://arxiv.org/abs/1908.00005}{{\tt
  1908.00005}}].

\bibitem{Campbell:2018wfu}
J.~M. Campbell, J.~Rojo, E.~Slade and C.~Williams, \emph{{Direct photon
  production and PDF fits reloaded}},
  \href{http://dx.doi.org/https://10.1140/epjc/s10052-018-5944-4}{\emph{Eur.
  Phys. J. C} {\bf 78} (2018) 470},
  [\href{http://arxiv.org/abs/1802.03021}{{\tt 1802.03021}}].

\bibitem{Bertone:2017bme}
{\scshape NNPDF} collaboration, V.~Bertone, S.~Carrazza, N.~P. Hartland and
  J.~Rojo, \emph{{Illuminating the photon content of the proton within a global
  PDF analysis}},
  \href{http://dx.doi.org/https://10.21468/SciPostPhys.5.1.008}{\emph{SciPost
  Phys.} {\bf 5} (2018) 008}, [\href{http://arxiv.org/abs/1712.07053}{{\tt
  1712.07053}}].

\bibitem{Manohar:2016nzj}
A.~Manohar, P.~Nason, G.~P. Salam and G.~Zanderighi, \emph{{How bright is the
  proton? A precise determination of the photon parton distribution function}},
  \href{http://dx.doi.org/https://10.1103/PhysRevLett.117.242002}{\emph{Phys.
  Rev. Lett.} {\bf 117} (2016) 242002},
  [\href{http://arxiv.org/abs/1607.04266}{{\tt 1607.04266}}].

\bibitem{Manohar:2017eqh}
A.~V. Manohar, P.~Nason, G.~P. Salam and G.~Zanderighi, \emph{{The Photon
  Content of the Proton}},
  \href{http://dx.doi.org/https://10.1007/JHEP12(2017)046}{\emph{JHEP} {\bf 12}
  (2017) 046}, [\href{http://arxiv.org/abs/1708.01256}{{\tt 1708.01256}}].

\bibitem{deFlorian:2014xna}
D.~de~Florian, R.~Sassot, M.~Epele, R.~J. Hern\'andez-Pinto and M.~Stratmann,
  \emph{{Parton-to-Pion Fragmentation Reloaded}},
  \href{http://dx.doi.org/https://10.1103/PhysRevD.91.014035}{\emph{Phys. Rev.
  D} {\bf 91} (2015) 014035}, [\href{http://arxiv.org/abs/1410.6027}{{\tt
  1410.6027}}].

\bibitem{deFlorian:2007ekg}
D.~de~Florian, R.~Sassot and M.~Stratmann, \emph{{Global analysis of
  fragmentation functions for protons and charged hadrons}},
  \href{http://dx.doi.org/https://10.1103/PhysRevD.76.074033}{\emph{Phys. Rev.
  D} {\bf 76} (2007) 074033}, [\href{http://arxiv.org/abs/0707.1506}{{\tt
  0707.1506}}].

\bibitem{Pedregosa:2011ork}
F.~Pedregosa et~al., \emph{{Scikit-learn: Machine Learning in Python}},
  \href{http://dx.doi.org/https://doi.org/10.48550/arXiv.1201.0490}{\emph{J.
  Machine Learning Res.} {\bf 12} (2011) 2825--2830},
  [\href{http://arxiv.org/abs/1201.0490}{{\tt 1201.0490}}].

\bibitem{Ochoa-Oregon:2023ktx}
S.~A. Ochoa-Oregon, D.~F. Renter\'\i{}a-Estrada, R.~J. Hern\'andez-Pinto and
  G.~F.~R. Sborlini, \emph{{Constraining fragmentation functions through
  hadron-photon production at higher-orders}},
  \href{http://dx.doi.org/https://10.1103/PhysRevD.107.096002}{\emph{Phys. Rev.
  D} {\bf 107} (2023) 096002}, [\href{http://arxiv.org/abs/2303.04965}{{\tt
  2303.04965}}].

\end{thebibliography}\endgroup

\end{document}